\documentclass{iopart}
\usepackage{iopams}
\usepackage{graphicx}

\begin{document}

\title[Saari's homographic conjecture]{Saari's homographic conjecture
	for planar equal-mass three-body problem
	under a strong force potential}

\author{
Toshiaki Fujiwara$^1$,
Hiroshi Fukuda$^2$,
Hiroshi Ozaki$^3$\\
and Tetsuya Taniguchi$^4$
}
\address{
$^{1,2,4}$ College of Liberal Arts and Sciences, Kitasato University,
1-15-1 Kitasato, Sagamihara, Kanagawa 252-0329, Japan
}
\address{
$^3$ General Education Program Center, Tokai University, Shimizu Campus,
3-20-1, Orido, Shimizu, Shizuoka 424-8610, Japan
}

\eads{
$^{1}$fujiwara@kitasato-u.ac.jp,
$^{2}$fukuda@kitasato-u.ac.jp,
$^{3}$ozaki@tokai-u.jp,
$^{4}$tetsuya@kitasato-u.ac.jp
}

\begin{abstract}
Donald Saari conjectured that the $N$-body motion
with constant configurational measure
is a motion with fixed shape. 
Here,
the configurational measure $\mu$ is a 
scale invariant product of
the moment of inertia $I=\sum_k m_k |q_k|^2$
and the potential function $U=\sum_{i<j} m_i m_j/|q_i-q_j|^\alpha$,
$\alpha >0$.
Namely, $\mu = I^{\alpha/2}U$.
We will show that
this conjecture is true
for planar equal-mass three-body problem 
under the strong force potential  $\sum_{i<j} 1/|q_i-q_j|^2$.
\end{abstract}

\submitto{\JPA}

\maketitle

\section{Saari's homographic conjecture}
\label{Introduction}
In 1969, Donald Saari conjectured that
if a $N$-body system has a constant moment of inertia
then the motion is a rotation with constant mutual distances $r_{ij}$ \cite{SaariOriginal}.
Here, the moment of inertia $I$ is defined by
\begin{equation}
I = \sum_k m_k |q_k|^2,
\end{equation}
with $m_k$ and $q_k$ being the masses and position vectors of body $k=1,2,3,\dots N$.
This is now called Saari's original conjecture.
In the conference ``Saarifest 2005'' at Guanajuato Mexico,
Richard Moeckel proved that the original conjecture is true for three-body problem
in $\mathbb{R}^d$ for any $d \ge 2$ \cite{MockelProof,MoeckelSaarifest}.

In the same conference,
Saari extended his conjecture. 
His new conjecture is 
``if the configurational measure $I^{\alpha/2} U$ is constant
then the $N$-body motion is homographic" \cite{SaariCollisions,SaariSaarifest},
where, 
\begin{equation}
U=\sum_{i<j} \frac{m_i m_j}{r_{ij}^\alpha}
\label{def:U}
\end{equation}
is the potential function.
This is indeed a natural extension of the original conjecture.
%
Note that a solution $q_k$ of $N$ bodies is called 
{\it homographic} 
if the configuration formed by the $N$ bodies moves 
in such a way as to remain similar to itself. 
For $\alpha\ne 2$, we can show that 
if the moment of inertia is constant then $U$ is constant, 
therefore the configurational measure is also constant.
For $\alpha = 2$, on the other hand,
$I=$ constant does not yields $U=$ constant \cite{Chenciner1997}.
Actually, there are some counter examples for the original conjecture for $\alpha=2$
\cite{Chenciner2002}\cite{Roberts}.
However, the extended conjecture is expected to be true for $\alpha=2$
and all $\alpha >0$.

Florin Diacu, Toshiaki Fujiwara, Ernesto P\'erez-Chavela and Manuele Santoprete
called this conjecture the ``Saari's homographic conjecture" 
and partly proved this conjecture for some cases  \cite{DiacHomographic}.
No one proved this conjecture completely, as far as we know.

Obviously,
Saari's conjecture is related to the motion in shape.
Here, a shape is a configuration of $N$ bodies up to rotation and scaling. 
To prove the Saari's original  and homographic conjecture,
it is important to find appropriate variables to describe motion in shape.
The moment of inertia $I$ describes the motion in size,
and the angular momentum $C$ describes the rotation.
What are the appropriate variables to describe the motion in shape?

An answer was given by Richard Moeckel and Richard Montgomery \cite{M&M}.
They used the ratio of the Jacobi coordinate to describe the motion in shape
for planar three-body problem.
Let us explain precisely.
To avoid non-essential complexity, let us consider equal masses case,
and set $m_k=1$ in this paper.
We take the center of mass frame. So, we have
\begin{equation}
\sum_k q_k =0.
\label{CM0}
\end{equation}
In the three-body problem, we have two Jacobi coordinates,
\begin{eqnarray}
z_1&=q_2-q_1,\\
z_2&=q_3-\frac{q_1+q_2}{2}=\frac{3}{2}q_3.
\end{eqnarray}
Since, we are considering  planar motions,
let us identify $q_k$ and $z_i$ with complex numbers.
Then, we can define the ratio of the Jacobi coordinates,
\begin{equation}
\label{eq:zeta}
\zeta = \frac{z_2}{z_1}=\frac{3}{2}\frac{q_3}{q_2-q_1}.
\end{equation}
Note that the variable $\zeta$ is invariant under the size change and rotation,
$q_k \mapsto \lambda e^{i\theta} q_k$ with $\lambda, \theta \in \mathbb{R}$.
Therefore, $\zeta$ depends only on  the shape.
The great idea by Moeckel and Montgomery is to use the variable $\zeta$
to describe the shape.
They actually write down the Lagrangian by 
the  variable $\zeta$, the moment of inertia $I$ and the rotation angle $\theta$.
They also write down the equations of motion for these variables.

Using the formulation developed by Moeckel and Montgomery,
we will show that the Saari's homographic conjecture is true
for planar equal-mass three-body problem under the strong force potential,
\begin{equation}
\label{eq:U}
U = \sum_{i<j} \frac{1}{|q_i-q_j|^2}.
\end{equation}
Namely, we will show that $d\zeta/dt=0$ if and only if $I U=$ constant.

In the section \ref{Lagrangian}, we derive the Lagrangian 
in terms of $I$, $\theta$ and $\zeta$
by elementary calculations.
The equations of motion and some useful relations are also shown in this section.
Every relations in the section \ref{Lagrangian} are valid for $\alpha \ne 0$.
In the section \ref{Proof},
we concentrate on the strong force potential $\alpha=2$.
We will  prove the Saari's homographic conjecture for this case.
Details in calculation are shown in  \ref{Mathematica}.
In the  \ref{properties},
some properties of the shape variable $\zeta$,
which may useful to understand this variable.

\section{Lagrangian for planar equal-mass three-body problem
in terms of shape, size and rotation angle}
\label{Lagrangian}
In this section,
we consider the planar equal-mass three-body problem under the
potential function 
(\ref{def:U}) with $m_1=m_2=m_3=1$ for
$\alpha \ne 0$.
Let $K=\sum_k \left| dq_k/dt \right|^2$ be  twice of the kinetic energy, 
and let the Lagrangian and the total energy be
$L=K/2+U/\alpha$ and
$E = K/2-U/\alpha$, respectively.
%
In the center of mass frame (\ref{CM0})
%
all quantities $\xi_k=q_k/(q_2-q_1)$ are expressed by the shape variable
in (\ref{eq:zeta}), as follows,
\begin{eqnarray}
\xi_1=\frac{q_1}{q_2-q_1}&=-\frac{1}{2}-\frac{\zeta}{3}, \label{eq:xi1}\\
\xi_2=\frac{q_2}{q_2-q_1}&=+\frac{1}{2}-\frac{\zeta}{3}, \label{eq:xi2}\\
\xi_3=\frac{q_3}{q_2-q_1}&=\frac{2}{3}\zeta. \label{eq:xi3}
\end{eqnarray}
Obviously, the triangle made of $q_1,q_2,q_3$ is similar to 
the triangle made of $\xi_1, \xi_2, \xi_3$.
Therefore, there are some $I \ge 0$ and $\theta \in \mathbb{R}$, such that
\begin{equation}
\label{eq:q}
q_k = \sqrt{I}\ e^{i\theta} \frac{\xi_k}{\sqrt{\sum_l |\xi_\ell|^2}}.
\end{equation}
We treat $I$, $\theta$ and $\zeta$ as independent  dynamical variables.

\subsection{Lagrangian}
Then, direct calculations for $K$ yields
\begin{eqnarray}
K
&=\sum_k \left|
	\frac{\dot I}{2\sqrt{I}}\ \frac{\xi_k}{\sqrt{\sum_l |\xi_\ell|^2}}
	+i\dot\theta \sqrt{I}\ \frac{\xi_k}{\sqrt{\sum_l |\xi_\ell|^2}}
	+\sqrt{I}\ \frac{d}{dt}\left( \frac{\xi_k}{\sqrt{\sum_l |\xi_\ell|^2}} \right)
	\right|^2
	\nonumber\\
%
&=\frac{\dot{I}^2}{4I}
	+ I \left(
		\dot\theta +\frac{\frac{2}{3}\zeta\wedge \dot\zeta}{\frac{1}{2}+\frac{2}{3}|\zeta|^2}
	\right)^2
	+\frac{I}{3}\frac{|\dot\zeta|^2}{\left( \frac{1}{2}+\frac{2}{3}|\zeta|^2 \right)^2}. \label{eq:calcK}
\end{eqnarray}
Here, 
the wedge product $\wedge$ represents
$(a+ib)\wedge(c+id)=ad-bc$
for  $a,b,c,d \in \mathbb{R}$, 
and the dot $d/dt$. 
On the other hand, the potential function $U$ is
\begin{eqnarray}
U
&=\frac{1}{I^{\alpha/2}}\left( \sum_k |\xi_k|^2\right)^{\alpha/2}
	\sum_{i<j} \frac{1}{|\xi_i-\xi_j|^{\alpha}}\nonumber\\
&=\frac{1}{I^{\alpha/2}}
	\left( \frac{1}{2}+\frac{2}{3}|\zeta|^2\right)^{\alpha/2}
	\left( 1+\frac{1}{|\zeta-1/2|^\alpha} + \frac{1}{|\zeta+1/2|^\alpha} \right).
\end{eqnarray}
Therefore, the configurational measure $\mu$ is a function of the shape variable $\zeta$,
\begin{equation}
\mu(\zeta)
=I^{\alpha/2}U
=\left( \frac{1}{2}+\frac{2}{3}|\zeta|^2\right)^{\alpha/2}
	\left( 1+\frac{1}{|\zeta-1/2|^\alpha} + \frac{1}{|\zeta+1/2|^\alpha} \right).
\end{equation}
Thus, we get the Lagrangian
\begin{equation}
L
=\frac{\dot{I}^2}{8I}
	+ \frac{I}{2} \left(
		\dot\theta+\frac{\frac{2}{3}\zeta\wedge \dot\zeta}{\frac{1}{2}+\frac{2}{3}|\zeta|^2}
	\right)^2
	+\frac{I}{6}\frac{|\dot\zeta|^2}{\left( \frac{1}{2}+\frac{2}{3}|\zeta|^2 \right)^2}
	+\frac{\mu(\zeta)}{\alpha I^{\alpha/2}},
\end{equation}
in terms of $I$, $\theta$, $\zeta$ and their velocities.
Or, identifying $\zeta=x+iy$ with $x,y \in \mathbb{R}$
to a two dimensional vector $\mathbf{x}=(x,y)$,
the Lagrangian is expressed as
\begin{equation}
L
=\frac{\dot{I}^2}{8I}
	+ \frac{I}{2} \left(
		\dot\theta
		+\frac{\frac{2}{3}\mathbf{x}\wedge\dot{\mathbf{x}}}{\frac{1}{2}+\frac{2}{3}|\mathbf{x}|^2}
	\right)^2
	+\frac{I}{6}\frac{|\dot{\mathbf{x}}|^2}{\left( \frac{1}{2}+\frac{2}{3}|\mathbf{x}|^2 \right)^2}
	+\frac{\mu(\mathbf{x})}{\alpha I^{\alpha/2}},
\end{equation}
with $\mathbf{x}\wedge\dot{\mathbf{x}}=x\dot y-y\dot x$ and
\begin{equation}
\fl
\mu(\mathbf{x})
=\left( \frac{1}{2}+\frac{2}{3}|\mathbf{x}|^2\right)^{\alpha/2}
	\left( 
		1
		+\frac{1}{\big((x-1/2)^2+y^2\big)^{\alpha/2}} 
		+\frac{1}{\big((x+1/2)^2+y^2\big)^{\alpha/2}} 
	\right).
\end{equation}

\subsection{Equation of motion for rotation angle} 
Obviously, the variable $\theta$ is cyclic. Therefore, we get the conservation law
of the angular momentum,
\begin{equation}
\label{eq:angularMomentum}
C = \frac{\partial L}{\partial \dot{\theta}}
=I \left(
		\dot\theta
		+\frac{\frac{2}{3}\mathbf{x}\wedge\dot{\mathbf{x}}}{\frac{1}{2}+\frac{2}{3}|\mathbf{x}|^2}
	\right)
= \mbox{ constant}.
\end{equation}
%
Then, the kinetic energy $K/2$ is given by
\begin{equation}
\label{eq:PartsOfKineticEnergy}
\frac{K}{2}
=\frac{\dot{I}^2}{8I}
	+ \frac{C^2}{2I}
	+\frac{I}{6}\frac{|\dot{\mathbf{x}}|^2}{\left( \frac{1}{2}+\frac{2}{3}|\mathbf{x}|^2 \right)^2}.
\end{equation}
The three terms in the right hand side represent
kinetic energies 
for the scale change,
for the rotation and for the shape change.

\subsection{Equation of motion for the moment of inertia}
The Euler-Lagrange  equation for the moment of inertia $I$,
\begin{equation}
\frac{d}{dt}\left(\frac{\partial L}{\partial \dot{I}}\right)-\frac{\partial L}{\partial I}=0
\end{equation}
yields
\begin{eqnarray}
\label{eq:I1}
\frac{\ddot I}{4}
&=\frac{\dot{I}^2}{8I}
	+ \frac{C^2}{2I}
	+\frac{I}{6}\frac{|\dot{\mathbf{x}}|^2}{\left( \frac{1}{2}+\frac{2}{3}|\mathbf{x}|^2 \right)^2}
	-\frac{1}{2}\frac{\mu(\mathbf{x})}{I^{\alpha/2}}\\
&=E + \left( \frac{1}{\alpha}-\frac{1}{2}\right) U.\label{eq:I2}
\end{eqnarray}

Multiplying $\dot{I}$ both side of the equation (\ref{eq:I2}),
we get
\begin{equation}
-\frac{\mu}{\alpha}(1-\alpha/2)I^{-\alpha/2}\dot{I}
=E\dot{I} - \frac{1}{4}\dot{I}\ddot{I}.
\end{equation}
This means 
\begin{eqnarray}
-\frac{\mu}{\alpha}\frac{d}{dt}\left( I^{1-\alpha/2} \right)
&=\frac{d}{dt}\left( EI - \frac{\dot{I}^2}{8} \right)\\
&=\frac{d}{dt}\left( 
		\frac{C^2}{2}
		+\frac{I^2}{6}\frac{|\dot{\mathbf{x}}|^2}{(\frac{1}{2}+\frac{2}{3}|\mathbf{x}|^2)^2}
		-\frac{\mu}{\alpha}I^{1-\alpha/2}
		\right).
\end{eqnarray}
Therefore, we get
\begin{equation}
\frac{d}{dt}\left( 
		\frac{I^2}{6}\frac{|\dot{\mathbf{x}}|^2}{(\frac{1}{2}+\frac{2}{3}|\mathbf{x}|^2)^2}
		\right)
=\frac{I^{1-\alpha/2}}{\alpha}\frac{d\mu}{dt}. \label{eq:SaariRelation}
\end{equation}
This relation was first derived by Saari \cite{SaariCollisions}.
We would like to call this ``Saari's relation''.
Inspired by this relation, let us introduce new `time' variable $s$ defined by
\begin{equation}
ds=\frac{1}{I} \left( \frac{1}{2}+\frac{2}{3}|\mathbf{x}|^2\right) dt.
\end{equation}
Then, we have
\begin{equation}
\frac{d}{dt} = \frac{1}{I}\left( \frac{1}{2}+\frac{2}{3}|\mathbf{x}|^2\right) \frac{d}{ds}
\end{equation}
and the Saari's relation (\ref{eq:SaariRelation}) is
\begin{equation}
\label{eq:SaariRelationInS}
\frac{d}{ds}\left( \frac{1}{6}\left|\frac{d\mathbf{x}}{ds}\right|^2 \right)
=\frac{I^{1-\alpha/2}}{\alpha}\frac{d\mu}{ds}.
\end{equation}

\subsection{Equation of motion for the shape variables}
The Euler-Lagrange equation  for $\mathbf{x}$,
\begin{equation}
\frac{d}{dt}\left( \frac{\partial L}{\partial \dot{\mathbf{x}}} \right)
	-\frac{\partial L}{\partial \mathbf{x}}
=0,
\end{equation}
yields
\begin{eqnarray}
\fl
\frac{d}{dt}\Bigg(
		\frac{I}{3(1/2+2|\mathbf{x}|^2/3)^2}\frac{d\mathbf{x}}{dt}
		-\frac{2C}{3}\frac{1}{(1/2+2|\mathbf{x}|^2/3)}(y,-x)
	\Bigg)\nonumber\\
\fl
=\frac{2C}{3}\frac{1}{(1/2+2|\mathbf{x}|^2/3)}\frac{d}{dt}(y,-x)
	+\frac{2C}{3}
		\left(\mathbf{x}\wedge\frac{d\mathbf{x}}{dt}\right)
		\frac{\partial}{\partial \mathbf{x}} \left( \frac{1}{(1/2+2|\mathbf{x}|^2/3)}\right)\nonumber\\
	+\frac{I}{6}\left| \frac{d\mathbf{x}}{dt}\right|^2 
		\frac{\partial}{\partial \mathbf{x}}\left( \frac{1}{(1/2+2|\mathbf{x}|^2/3)^2}\right)
	+\frac{1}{\alpha I^{\alpha/2}}\frac{\partial \mu}{\partial \mathbf{x}}.
\end{eqnarray}
Using the `time' variable $s$, this equation of motion is
\begin{equation}
\label{eq:equationOfMotionBys}
\frac{d^2 \mathbf{x}}{ds^2}
=\frac{\displaystyle{2C-\frac{4}{3}\left( \mathbf{x}\wedge\frac{d\mathbf{x}}{ds}\right)}}
	{\displaystyle{\frac{1}{2}+\frac{2}{3}|\mathbf{x}|^2}}
	\left(
		\frac{dy}{ds},-\frac{dx}{ds}
	\right)
	+\frac{3I^{1-\alpha/2}}{\alpha}\frac{\partial \mu}{\partial \mathbf{x}}.
\end{equation}
Inner product of $d\mathbf{x}/ds$ and $d^2 \mathbf{x}/ds^2$  yields
\begin{equation}
\frac{d\mathbf{x}}{ds}\cdot\frac{d^2 \mathbf{x}}{ds^2}
=\frac{3 I^{1-\alpha/2}}{\alpha} \frac{d\mathbf{x}}{ds}\cdot\frac{\partial \mu}{\partial \mathbf{x}}
=\frac{3 I^{1-\alpha/2}}{\alpha} \frac{d\mu}{ds}.
\end{equation}
This is nothing but the Saari's relation in (\ref{eq:SaariRelationInS}).
While, the wedge product of the same pair yields
\begin{equation}
\label{eq:fromTheEquationOfMotion}
\frac{d\mathbf{x}}{ds}\wedge\frac{d^2 \mathbf{x}}{ds^2}
=-\frac{\displaystyle{2C-\frac{4}{3}\left( \mathbf{x}\wedge\frac{d\mathbf{x}}{ds}\right)}}
	{\displaystyle{\frac{1}{2}+\frac{2}{3}|\mathbf{x}|^2}}
	\left| \frac{d\mathbf{x}}{ds} \right|^2
	+\frac{3I^{1-\alpha/2}}{\alpha} 
		\frac{d\mathbf{x}}{ds}\wedge\frac{\partial \mu}{\partial \mathbf{x}}.
\end{equation}
Every equations are valid for all $\alpha\ne 0$.
We will use these expressions later.

\section{Proof of the Saari's homographic conjecture under the strong force potential}
\label{Proof}
In this section,
we will prove the Saari's homographic conjecture for the case $\alpha=2$,
namely,
$\dot{\zeta}=0$ if and only if $\mu=$ constant .

Let us assume
\begin{equation}
\mu=\mu_0=\mbox{ constant}.
\end{equation}
Then, by the Saari's relation (\ref{eq:SaariRelationInS}), we have
\begin{equation}
\label{eq:constantVelocity}
\left| \frac{d\mathbf{x}}{ds}\right|^2 = k^2
\end{equation}
with constant $k\ge 0$.

If $k=0$, then $d\mathbf{x}/ds=0$, namely $\dot{\zeta}=0$.

\begin{figure} 
   \centering
   \includegraphics[width=7cm]{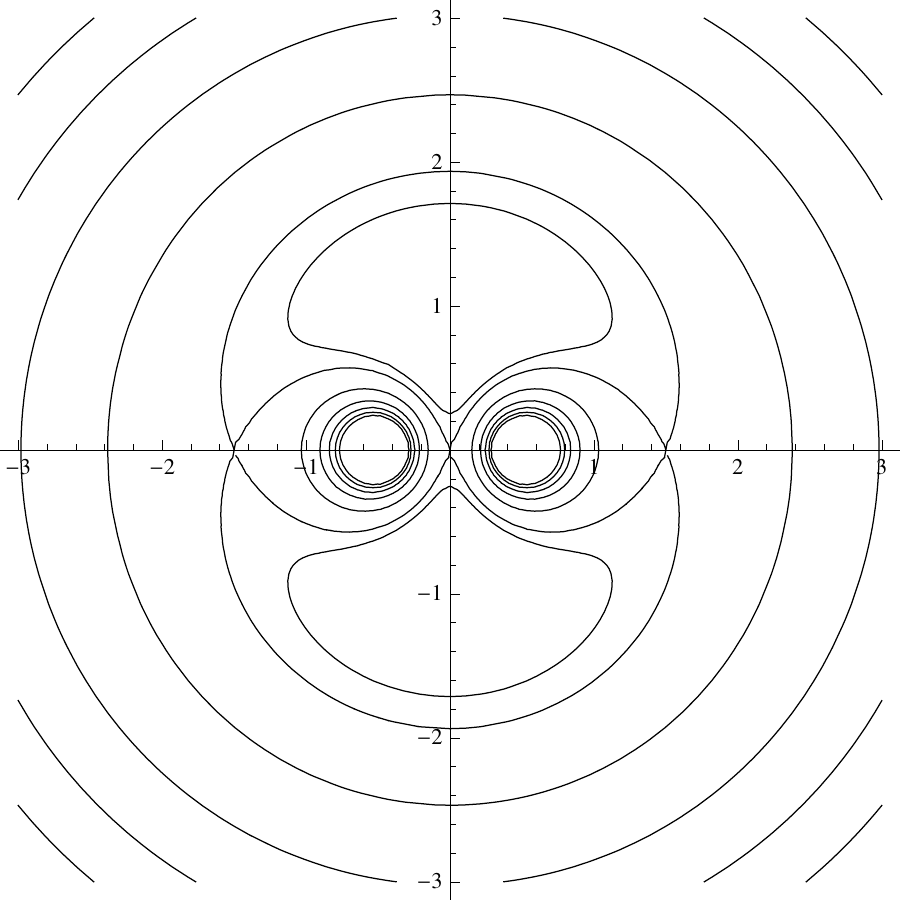} 
   \caption{Curves of $\mu(x,y)=\mu_0$ for several $\mu_0$.
}
   \label{fig:contourForMu}
\end{figure}

Let us examine the case $k>0$.
For this case, 
the point $\mathbf{x}(s)$ moves on the curve $\mu(\mathbf{x})=\mu_0$ with constant speed $|d\mathbf{x}/ds|=k$.
This motion of $\mathbf{x}(s)$ is not able to keep 
$\partial \mu/\partial \mathbf{x}=0$.
Because, the points that satisfy $\partial \mu/\partial \mathbf{x}=0$ are
only five central configurations at
$\mathbf{x} =(\pm 3/2,0), (0,0), (0,\pm \sqrt{3}/{2})$. 
See figure \ref{fig:contourForMu}.

So, we can take some finite arc of the curve $\mu(\mathbf{x})=\mu_0$ on which
\begin{equation}
\frac{\partial \mu}{\partial \mathbf{x}}\ne 0.
\end{equation}
Then the equation (\ref{eq:constantVelocity}) and 
\begin{equation}
\frac{d\mathbf{x}}{ds}\cdot \frac{\partial \mu}{\partial \mathbf{x}} = \frac{d\mu}{ds}=0
\end{equation}
yields
\begin{equation}
\label{eq:velocity}
\frac{d\mathbf{x}}{ds} 
= \frac{\epsilon k}{|\partial \mu/\partial \mathbf{x}|} 
	\left(
		-\frac{\partial \mu}{\partial y}, \frac{\partial \mu}{\partial x}
	\right).
\end{equation}
Here, $\epsilon = \pm 1$ determines the direction of the motion on the curve $\mu=\mu_0$.
Differentiate this expression by $s$ again, we have
\begin{equation}
\frac{d^2\mathbf{x}}{ds^2}
=\frac{\epsilon k}{|\partial \mu/\partial \mathbf{x}|}
	\frac{d}{ds}\left( -\frac{\partial \mu}{\partial y},\frac{\partial \mu}{\partial x}\right)
	+\left( -\frac{\partial \mu}{\partial y},\frac{\partial \mu}{\partial x}\right)
		\frac{d}{ds}\left( \frac{\epsilon k}{|\partial \mu/\partial \mathbf{x}|} \right).
\end{equation}
Therefore, we get
\begin{eqnarray}
&\frac{d\mathbf{x}}{ds}\wedge\frac{d^2\mathbf{x}}{ds^2}\nonumber\\
&=\frac{k^2}{|\partial \mu/\partial \mathbf{x}|^2}
	\left( -\frac{\partial \mu}{\partial y},\frac{\partial \mu}{\partial x}\right)
	\wedge
	\frac{d}{ds}
	\left( -\frac{\partial \mu}{\partial y},\frac{\partial \mu}{\partial x}\right)\nonumber\\
&=\frac{\epsilon k^3}{|\partial \mu/\partial \mathbf{x}|^3}
	\Bigg(
		\left(\frac{\partial\mu}{\partial y}\right)^2 \frac{\partial^2 \mu}{\partial x^2}
		-2 \frac{\partial \mu}{\partial x}\frac{\partial \mu}{\partial y} \frac{\partial^2 \mu}{\partial x \partial y}
		+\left(\frac{\partial\mu}{\partial x}\right)^2 \frac{\partial^2 \mu}{\partial y^2}
	\Bigg).
\end{eqnarray}
Therefore, the curvature $\rho^{-1}$ of the curve $\mu=\mu_0$ should be
\begin{equation}
\label{eq:kappa}
\rho^{-1}
=\frac{\epsilon}{|\partial \mu/\partial \mathbf{x}|^3}
	\Bigg(
		\left(\frac{\partial\mu}{\partial y}\right)^2 \frac{\partial^2 \mu}{\partial x^2}
		-2 \frac{\partial \mu}{\partial x}\frac{\partial \mu}{\partial y} \frac{\partial^2 \mu}{\partial x \partial y}
		+\left(\frac{\partial\mu}{\partial x}\right)^2 \frac{\partial^2 \mu}{\partial y^2}
	\Bigg).
\end{equation}

On the other hand,
by the relation (\ref{eq:fromTheEquationOfMotion}) which is a result of the equation of motion
and the expression for the velocity (\ref{eq:velocity}),
we have another expression for the curvature
\begin{equation}
\label{eq:kappa2}
\rho^{-1}
=\frac{1}{1/2+2|\mathbf{x}|^2/3}
	\left(
		-\frac{2C}{k}
		+\frac{4\epsilon}{3|\partial \mu/\partial \mathbf{x}|}
			\mathbf{x}\cdot\frac{\partial \mu}{\partial \mathbf{x}}
	\right)
	-
	\frac{3\epsilon}{2k^2}\left|\frac{\partial \mu}{\partial \mathbf{x}}\right|.
\end{equation} 

Our plan to exclude the case $k>0$ is the following.
Since $|d\mathbf{x}/ds|=k=$ constant,
the parameter $s$ is proportional to the arc length of the curve $\mathbf{x}(s)$
on $\mu=\mu_0$.
Therefore, if $k>0$ there must be finite arc,
on which the  curvature (\ref{eq:kappa}) coincides with (\ref{eq:kappa2}).
We  call such arc non-Saari arc.
%
In the following, we will show that non-Saari arc does not exist.
Namely, $k>0$ is impossible.

Let us examine the condition for the two curvature have the same value.
The condition is
\begin{eqnarray}
\fl
\frac{1}{|\partial \mu/\partial \mathbf{x}|^3}
	\Bigg(
		\left(\frac{\partial\mu}{\partial y}\right)^2 \frac{\partial^2 \mu}{\partial x^2}
		-2 \frac{\partial \mu}{\partial x}\frac{\partial \mu}{\partial y} \frac{\partial^2 \mu}{\partial x \partial y}
		+\left(\frac{\partial\mu}{\partial x}\right)^2 \frac{\partial^2 \mu}{\partial y^2}
	\Bigg)\nonumber\\
=\frac{1}{1/2+2|\mathbf{x}|^2/3}
	\left(
		-\frac{2\epsilon C}{k}
		+\frac{4}{3|\partial \mu/\partial \mathbf{x}|}
			\mathbf{x}\cdot\frac{\partial \mu}{\partial \mathbf{x}}
	\right)
	-
	\frac{3}{2k^2}\left|\frac{\partial \mu}{\partial \mathbf{x}}\right|.
\end{eqnarray}
Therefore,
\begin{eqnarray}
\fl
\frac{4C^2}{k^2N^2}|\nabla \mu|^6
-\Bigg(
	\left(\frac{\partial\mu}{\partial y}\right)^2 \frac{\partial^2 \mu}{\partial x^2}
		-2 \frac{\partial \mu}{\partial x}\frac{\partial \mu}{\partial y} \frac{\partial^2 \mu}{\partial x \partial y}
		+\left(\frac{\partial\mu}{\partial x}\right)^2 \frac{\partial^2 \mu}{\partial y^2}
	-\frac{4}{3N}(\mathbf{x}\cdot \nabla \mu) |\nabla \mu|^2
	+\frac{3}{2k^2}|\nabla \mu|^4
	\Bigg)^2\nonumber\\
=0.
\end{eqnarray}
Where,
$N=1/2+2|\mathbf{x}|^2/3$ and $\nabla \mu=\partial \mu/\partial \mathbf{x}$.
The left hand side is a ratio of polynomials of $x^2$ , $y^2$, $C^2$ and $k^2$.
Let  the numerator of this ratio be 
a polynomial $P(x^2,y^2,C^2,k^2)$,
then $x^2$, $y^2$ must satisfy the following equation,
\begin{equation}
P(x^2,y^2,C^2,k^2)=0.
\end{equation}
The maximum power of the variables for $P$ are
$x^{60}$, $y^{60}$, $C^2$ and $k^4$. 

On the other hand,
the equation $\mu=\mu_0$ is also a ratio of polynomials of $x^2$, $y^2$ and $\mu_0$.
Let the numerator of this ratio be a polynomial $Q$, then we have the following equation
\begin{eqnarray}
Q(x^2,y^2,\mu_0)
=
&27+64 x^6+156 y^2+208 y^4+64 y^6\nonumber\\
&\ +48 x^4 \left(3+4 y^2\right)+4 x^2 \left(27+88 y^2+48 y^4\right)\nonumber\\
&\ -6 \mu_0 \left(16 x^4+8 x^2 \left(-1+4 y^2\right)+\left(1+4 y^2\right)^2\right)\nonumber\\
&=0.
\end{eqnarray}

The non-Saari arc must satisfy
both  $P=0$ and $Q=0$
for some value of parameters $C^2$, $k^2$ and $\mu_0$.

There is no finite arc with $x=x_0=$ fixed and $\mu=\mu_0$.
Because,
for $x=x_0$, $Q=0$ is a polynomial of $y^2$ of order $y^6$ with the coefficient
of $y^6$ being $64\ne 0$. Therefore, solutions of $y$ for $Q=0$ are discrete.
Thus, any finite arc must have some finite interval $x_1 \le x \le x_2$.

There is no finite interval $x_1 \le x \le x_2$
that every  $x$ in this interval satisfy $P=Q=0$.
To show this,
we eliminate $y^2$ from $P(x^2,y^2)=Q(x^2,y^2)=0$ to get
new polynomial $R(x^2)=0$.
This polynomial turns out to be order $x^{68}$,
\begin{equation}
\label{eq:R}
R =A\  x^8 (4x^2-1)^6\sum_{0\le n \le 24} c_n(C^2,k^2,\mu_0) x^{2n}.
\end{equation}
Where, $A$ is a big integer.
To have continuous solution of $x$ for $R=0$,
the polynomial $R$ must be identically equal to zero.
Namely, all coefficients $c_n$, $n=0,1,2,\dots,24$ must be zero.
Therefore, we have 25 conditions for only three parameters $C^2$, $k^2$ and $\mu_0$.
Actually, there are no parameters to make all 25 coefficients vanish.
See  \ref{Mathematica} for detail.
Therefore, there is no finite interval of $x$ on which $P=Q=0$ is satisfied.
Thus $k>0$ case is excluded.

Therefore,
we have proved that if $\mu=$ constant, then $\dot{\zeta}=0$,
namely the three-body keep its shape of the triangle being similar.

Inversely,
if $\dot{\zeta}=0$, then obviously $\mu(\zeta)$ is constant.
This completes a proof for the Saari's homographic conjecture
for the case $\alpha=2$
planar equal-mass three-body problem.

\ack
The authors thank
Richard Moeckel and Richard Montgomery
for  sending us their preprint
on the shape variable and the Lagrangian.
This research of one of the author T.~Fujiwara  has been  supported by
Grand-in-Aid for Scientific Research 23540249 JSPS.

\appendix
\section{Details in calculation}
\label{Mathematica}
In this Appendix, details in calculation are shown.
The following calculations were performed using Mathematica 8.0.1.0.

To eliminate the variable $y^2$ from the equation $P=Q=0$,
we calculate the resultant of $P$ and $Q$ with respect to $y^2$,
\begin{equation}
R=\mbox{Resultant}[P,Q,y^2].
\end{equation}
In the actual calculations, we replaced $y^2$ with $Y$
and calculated Resultant$[P,Q,Y]$, 
because Mathematica doesn't accept $y^2$ as a variable.

Then, the  coefficients $c_{n}$, $n=0,1,2,\dots,24$ in the equation (\ref{eq:R})
are polynomials of $C^2$, $k^2$ and $\mu_0$.
To eliminate $k^2$ from the equation $c_{24}=c_{23}=0$,
we again calculate the resultant
of $c_{24}$ and $c_{23}$ with respect to $k^2$,
\begin{eqnarray}
\fl
d_1
&=\mbox{Resultant}[c_{24},c_{23},k^2]\nonumber\\
\fl
&= D_1\ 
	\mu_0^6
	(\mu_0-1)^3
	(2\mu_0-1)^{12}
	\Bigg(
		(3\mu_0-1)C^2+2\mu_0(2\mu_0-1)^2
	\Bigg)
	C^2,
\end{eqnarray}
with a big  integer $D_1$.
The configurational measure $\mu$ is not smaller than  3,
\begin{eqnarray}
\fl
\mu
=\frac{1}{3}
	\left( \sum_{i<j} r_{ij}^2 \right)
	\left(\sum_{i<j} \frac{1}{r_{ij}^2}\right)
\ge 3 \left( r_{12}^2 r_{23}^2 r_{13}^2 \right)^{1/3}
	\left( \frac{1}{r_{12}^2 r_{23}^2 r_{13}^2} \right)^{1/3}
=3.
\end{eqnarray}
Therefore, $\mu_0 \ge 3$.
Then, $d_1=0$ yields
$C^2=0$.
For $C^2=0$,
the resultant of $c_{24}$ and $c_{22}$ with respect to $k^2$ is
\begin{eqnarray}
d_2
&=\mbox{Resultant}[c_{24},c_{22},k^2]\nonumber\\
&=D_2\ 
	\mu_0^{12}
	(\mu_0-1)^4
	(2\mu_0-1)^{16}
& \ne 0.
\end{eqnarray}
Where, $D_2$ is another big  integer.
Therefore, it is impossible to make 
$c_{24}=c_{23}=c_{22}=0$ simultaniously.

This completes a proof that there is no parameter $C^2$, $k^2$ and $\mu_0$
to make $R=0$ identically.

\section{Properties  of the shape variable}
\label{properties}
In this appendix, some properties of the shape variable $\zeta$ is shown.

\subsection{Geometrical interpretation  of the shape variable}
According to the work of Moeckel and Montgomery \cite{M&M},
we defined the shape variable $\zeta$ as the ratio of the Jacobi coordinates,
in the section \ref{Introduction}.
Here, we give a geometrical interpretation of this definition.

\begin{figure}
   \centering
   \includegraphics[width=4cm]{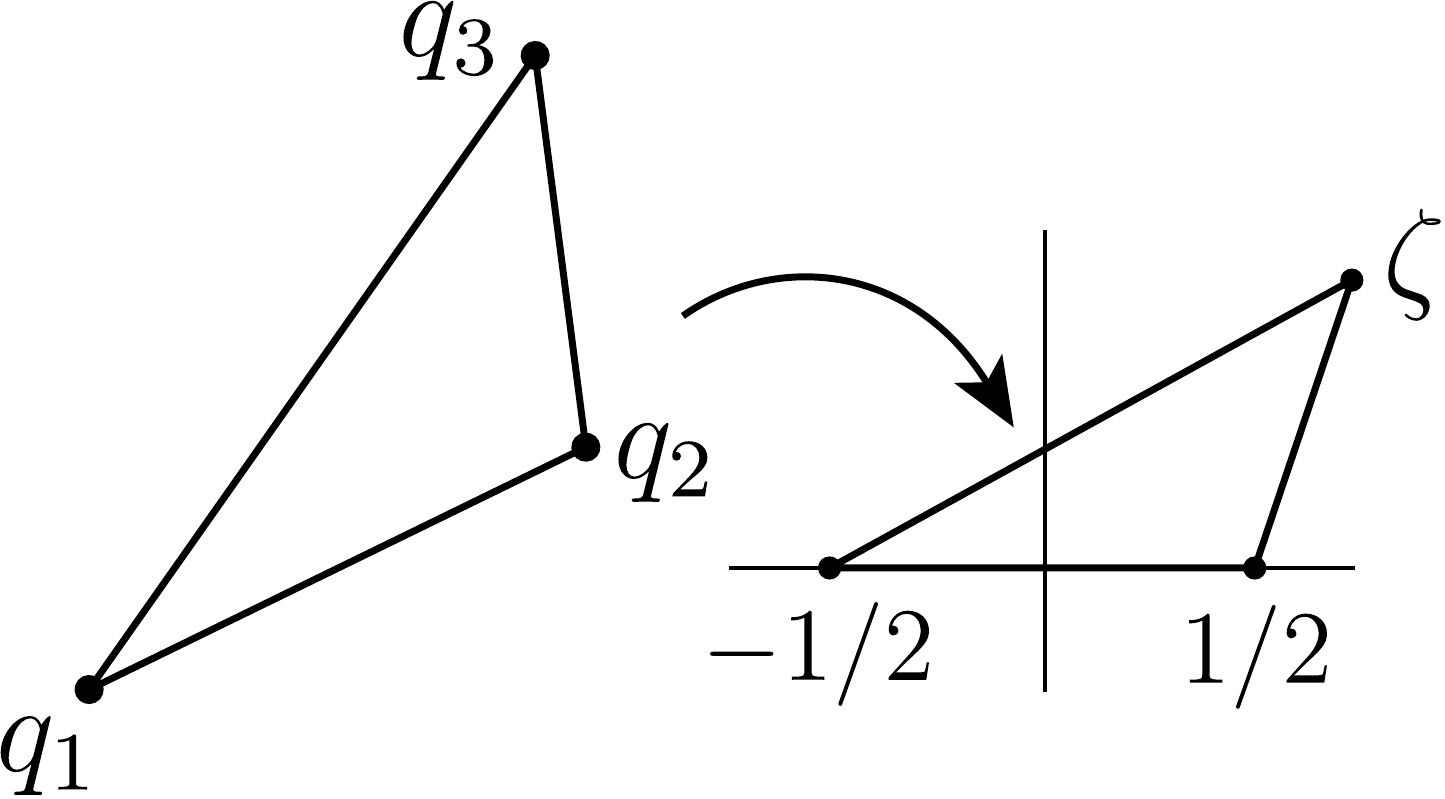} 
   \caption{The definition of shape variable $\zeta$.
   A transformation keeping similarity and the orientation
   transforms the triangle $q_1 q_2 q_3$ to the triangle $ab\zeta$
   with $a=-1/2$, $b=1/2$.}
   \label{fig:defOfZeta}
\end{figure}

For given triangle $q_{1}q_{2}q_{3}$,
we can transform the points $q_1 \mapsto a=-1/2$ 
and $q_2 \mapsto b=1/2$
keeping the similarity and the orientation of the triangle.
The points $a$ and $b$ are fixed.
Let $\zeta$ be the image of $q_3$ by this transformation.
So, the triangles $q_{1}q_{2}q_{3}$ and $ab\zeta$ are similar and
have same orientation. See figure \ref{fig:defOfZeta}.
It is clear that the variable $\zeta$ describe the shape of the triangle $q_{1}q_{2}q_{3}$.
This gives an alternative definition of the shape variable $\zeta$.

The center of mass of the triangle $ab\zeta$ is $\zeta/3$.
To make the center of mass being fixed to the origin,
we subtract $\zeta/3$ from the three vertices.
Thus, we have three vertices $\xi_1=-1/2-\zeta/3$, $\xi_2=1/2-\zeta/3$ and $\xi_3=2\zeta/3$.
These are the equations (\ref{eq:xi1})--(\ref{eq:xi3}).

\subsection{Contribution to the angular momentum of the shape change}
The motion in $\zeta$ contributes to
a size change and the angular momentum
as well as shape change.
To eliminate the contribution to the size change,
we normalized $\xi_k$ by the size $\sqrt{\sum_l |\xi_\ell|^2}$. 
Let us write the normalized variable $\eta_k$,
\begin{equation}
\eta_k = \frac{\xi_k}{\sqrt{\sum_l |\xi_\ell|^2}}.
\end{equation}
Then, $\eta_k$ have unit size,
therefore, have no contribution to the size change.

The variables $\eta_k$ still contribute to the angular momentum.
The equation (\ref{eq:angularMomentum})
describes that the total angular momentum 
is the sum of
the angular momentum of the  rotation angle $I\dot{\theta}$
and the angular momentum of the shape change $I \sum_k \eta_k\wedge \dot{\eta}_k$.

We can subtract the angular momentum from $\eta_k$.
The result is the variable $Q_k$, $k=1,2,3$ introduced by Diacu et al.
They defined $Q_k$ as follows,
\begin{equation}
Q_k(t)=\exp\left( -i \int_0^t \frac{Cdt}{I} \right) \frac{q_k}{\sqrt{I}}.
\end{equation}
They showed that $Q_k$ satisfy 
$\sum_k Q_k=0$, $\sum_k |Q_k|^2=1$ and $\sum_k Q_k \wedge \dot{Q}_k =0$ \cite{DiacHomographic}.
Namely, the variables $Q_k$ have unit size and no rotation (vanishing angular momentum).
They used  this variables $Q_k$ to analyze the shape change.

To have a expression for $Q_k$ by $\zeta$,
let us write the relation between $\zeta$ and $q_k$.
Integrating the equation (\ref{eq:angularMomentum}),
we get the expression for $\theta(t)$,
\begin{equation}
\theta(t)
=\int_0^t \frac{Cdt}{I}+\theta(0)
	-\frac{2}{3}\int_0^t \frac{\zeta \wedge \dot{\zeta} dt}{\frac{1}{2}+\frac{2}{3}|\zeta|^2}.
\end{equation}
Substituting this expression to the equation (\ref{eq:q}),
we get the relation,
\begin{equation}
q_k
=\sqrt{I}\exp
	\left(
		i\int_0^t \frac{Cdt}{I}
		+i\theta(0)
		-\frac{2i}{3}\int_0^t \frac{\zeta \wedge \dot{\zeta} dt}{\frac{1}{2}+\frac{2}{3}|\zeta|^2}
	\right)
    \frac{\xi_k}{\sqrt{\frac{1}{2}+\frac{2}{3}|\zeta|^2}},
\end{equation}
with the equations (\ref{eq:xi1})--(\ref{eq:xi3}) for $\xi_k$.
Now, we can express $Q_k$  by  $\zeta$,
\begin{equation}
Q_k
=\exp
	\left(
		i\theta(0)
		-\frac{2i}{3}\int_0^t \frac{\zeta \wedge \dot{\zeta} dt}{\frac{1}{2}+\frac{2}{3}|\zeta|^2}
	\right)
    \frac{\xi_k}{\sqrt{\frac{1}{2}+\frac{2}{3}|\zeta|^2}}.
    \label{eq:QandZeta}
\end{equation}
Note that the mutual relations between $q_k(t)$,  $Q_k(t)$ and $\zeta(t)$ are not local in time.
For example, the variable $Q_k(t)$ depend on the history of $\zeta(t')$ with $t \ge t' \ge 0$.

\subsection{Energy for the shape change}
The term ``energy for the shape change'' should be treated carefully.
In the equation (\ref{eq:PartsOfKineticEnergy}),
it is natural to understand the therm $\dot{I}^2/{(8I)}$ be the energy for the size change
and $C^2/{(2I)}$ be the energy for the rotaion.
The last term should be the ``energy for the shape change''.
This is not the kinetic energy of $\eta_k$, $I/2\ \sum_k |\dot{\eta_k}|^2$.
Because this term contains the rotation energy
$I/2 \ (\sum_k \eta_k \wedge \dot{\eta}_k)^2$,
which is counted in the energy for the rotation $C^2/(2I)$.
In the equation  (\ref{eq:calcK}),
the ``energy for the shape change''
is naturally defined by 
the kinetic energy minus the rotation energy,
\[
\frac{I}{2} \sum_k |\dot{\eta}_k|^2 
	- \frac{I}{2} \left(  \sum_k \eta_k \wedge \dot{\eta}_k  \right)^2
=\frac{I}{6}\frac{|\dot\zeta|^2}{\left( \frac{1}{2}+\frac{2}{3}|\zeta|^2 \right)^2}.
\]
One may be convinced of this result
by knowing that this is equal to $I/2\ \sum_k |\dot{Q}_k|^2$, the kinetic energy for $Q_k$
which have unit size and no rotation. 
To see this,
an alternative  expression for (\ref{eq:QandZeta})
\begin{equation}
Q_k
=\exp
	\left(
		i\theta(0)
		-i\int_0^t \sum_l \eta_\ell \wedge \dot{\eta}_\ell dt
	\right)
	\eta_k
\end{equation}
will be useful.

\end{document}